\documentclass[conference]{IEEEtran}
\IEEEoverridecommandlockouts
\usepackage{cite}
\usepackage{tabularx}
\usepackage{amsmath,amssymb,amsfonts}
\usepackage{algorithmic}
\usepackage{hyperref}
\usepackage{graphicx}
\usepackage{textcomp}
\usepackage{xcolor}
\usepackage{fancyhdr}
\usepackage{kantlipsum}
\fancypagestyle{plain}{
  \fancyhf{}
  \fancyhead[C]{Conference on \LaTeX}     
  \fancyfoot[L]{This is a notice}

}
\usepackage{eso-pic}

\def\BibTeX{{\rm B\kern-.05em{\sc i\kern-.025em b}\kern-.08em
    T\kern-.1667em\lower.7ex\hbox{E}\kern-.125emX}}

\begin{document}
\AddToShipoutPictureBG*{%
  \AtPageUpperLeft{%
    \setlength\unitlength{1in}%
    \hspace*{\dimexpr0.5\paperwidth\relax}
    \makebox(0,-0.75)[c]{\textit{Accepted for publication at the 10\textsuperscript{th} IEEE International Conference on Collaboration and Internet Computing (IEEE CIC 2024)}}
}}

\title{InsightPulse: An IoT-based System for User Experience Interview Analysis\\

\thanks{
This work was developed as part of the TECHIN-515: Hardware Software Laboratory 2 course project and supported by the Global Innovation Exchange at the University of Washington, Seattle, WA, USA.}
}

\author{
    \IEEEauthorblockN{
        Dian Lyu\textsuperscript{\S}\IEEEauthorrefmark{1}, 
        Yuetong Lu\textsuperscript{\S}\IEEEauthorrefmark{1}, 
        Jassie He\textsuperscript{\S}\IEEEauthorrefmark{1}, 
        Murad Mehrab Abrar\textsuperscript{\S}\IEEEauthorrefmark{2}, 
        Ruijun Xie\IEEEauthorrefmark{3}, 
        John Raiti\IEEEauthorrefmark{1}
    }
    \textsuperscript{\S}\textit{Co-first authors with equal contributions}\\
    \IEEEauthorblockA{
        \IEEEauthorrefmark{1}\textit{Global Innovation Exchange, University of Washington, Seattle, WA, USA}\\
        \IEEEauthorrefmark{2}\textit{Department of Mechanical Engineering, University of Washington, Seattle, WA, USA}\\
        \IEEEauthorrefmark{3}\textit{Department of Electrical Engineering, George Washington University, Washington, D.C., USA}
    }
    Email: \{dianlyu, yuetongl, yinghe, jraiti\}@uw.edu, mabrar@uw.edu, xrj0826@gmail.com
}

\maketitle

\begin{abstract}
Conducting efficient and effective user experience (UX) interviews often poses challenges, such as maintaining focus on key topics and managing the duration of interviews and post-interview analyses. To address these issues, this paper introduces InsightPulse, an Internet of Things (IoT)--based hardware and software system designed to streamline and enhance the UX interview process through speech analysis and Artificial Intelligence. InsightPulse provides real-time support during user interviews by automatically identifying and highlighting key discussion points, proactively suggesting follow-up questions, and generating thematic summaries. These features enable more insightful discoveries and help to manage interview duration effectively. Additionally, the system features a robust backend analytics dashboard that simplifies the post-interview review process, thus facilitating the quick extraction of actionable insights and enhancing overall UX research efficiency.

\end{abstract}

\vspace{10pt}


\section{Introduction}

Conducting effective user research is essential for developing user-centered designs and ensuring that products meet the needs and expectations of their target audiences \cite{Shluzas}. Direct interview is a common method of collecting useful data for user experience (UX) and requirements engineering (RE) \cite{Alvarez, Rohrer}. UX interview, in particular, is a critical component of the research process, as it provides valuable insights into user behavior, pain points, and desires. However, researchers frequently encounter several issues that can affect the quality of insights gathered during user interviews and extend the duration of interview sessions \cite{Wilson}. Major issues include: 

\begin{itemize}
    \item \textit{Loss of focus on key discussion topics:} Interviewers can easily diverge from critical themes, leading to less relevant data collection \cite{Cohene}.
    
    \item \textit{Inefficient management of interview duration:} Without proper tools, interviews can run longer than planned, causing participant fatigue and data quality degradation.
    
    \item \textit{Prolonged and cumbersome post-interview analyses:} Manual transcription and thematic analysis can be time consuming and error-prone, delaying actionable insights. 

\end{itemize}

To address these persistent issues, we introduce InsightPulse, a cost-effective IoT-based system that combines hardware and software to enhance the UX interview process. By integrating an AI-based speech analysis, the system provides real-time support during interviews and automates the key aspects of user research. InsightPulse is currently a project under development that includes a hardware device and a web application. We develop a dedicated hardware device to ensure a distraction-free environment for interviewers and maintain focus without being interrupted by apps and notifications that commonly occur when using smartphones or tablets to record interviews. The current version of the InsightPulse hardware device features:

\begin{itemize}
    \item \textit{Real-time Summary:} Automatically generates concise summary of the interview as it progresses and highlights the key points.

    \item \textit{Proactive Follow-up Question Suggestions:} Provides real-time suggestions with follow-up questions to help interviewers maintain focus.
    
    \item \textit{Timer:} Keeps track of the duration of the interview.
    
    \item \textit{Database Storage:} Stores interview conversation in text data and audio recording formats.

    \item \textit{Interview Privacy Protection:} Utilizes a Radio Frequency Identification (RFID)-based device activation to provide a layer of privacy to the interview recordings. The RFID card serves as an access control mechanism to ensure that only authorized individuals can operate the device.
    
\end{itemize}

The associated InsightPulse web application offers additional post-interview analysis functionalities, featuring:

\begin{itemize}
    \item \textit{Interview Review and Extraction of Actionable Insights:} Streamlines the process of reviewing interviews and extracting key insights.
    
    \item \textit{Identification and Highlighting of Key Discussion Points:} Ensures that critical themes are captured accurately and consistently throughout the interview.
    
    \item \textit{Generation of Thematic Summaries:} Provides immediate and structured insights to facilitate efficient data analysis.
\end{itemize}


The rest of the paper is organized as follows: Section II details the related work, Section III explains the methodology and hardware-software architecture, Section IV includes the cost and testing associated with InsightPulse, Section V outlines the limitations and future work, and Section VI concludes the paper.


\begin{figure*}[htbp]
\centerline{\includegraphics[width=15cm]{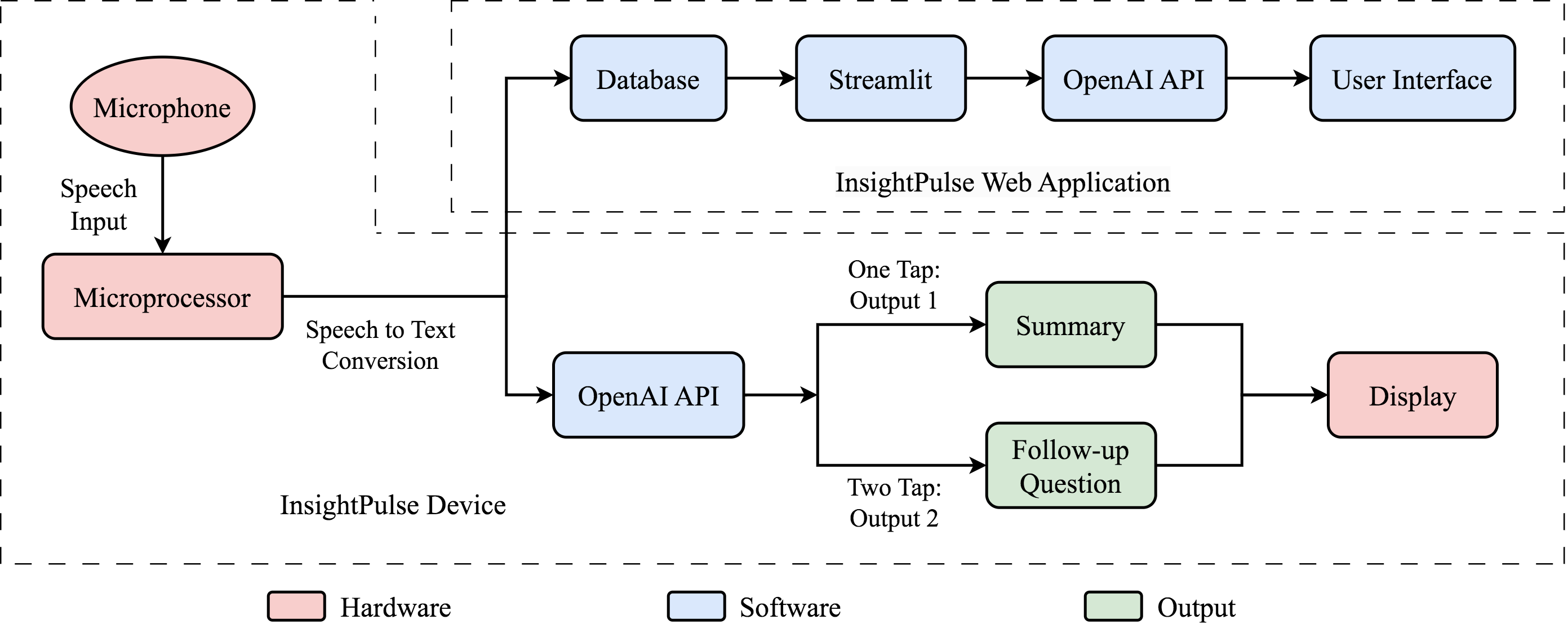}}
\caption{InsightPulse system architecture.}
\label{architecture}
\end{figure*}

\section{Related Work}

The development of InsightPulse is motivated by extensive research in the fields of AI-enabled UX research and explainable AI (XAI). This section reviews the related work that provided a foundation for understanding the technical and conceptual underpinnings of InsightPulse.

\subsection{AI-Enabled User Experience Research}

The integration of IoT and AI in UX research has been explored by several studies to deliver solutions with improved user experiences \cite{Lu, Yang}. In addition, the use of AI and machine learning capabilities for enhancing user experience is becoming a significant trend \cite{Abbas, Nikiforova}, creating numerous new opportunities for Human-Computer Interaction (HCI) and UX professionals \cite{Holmquist}. For instance, Yang et al. \cite{Yang-and-Bin} propose a methodology for using AI technology to measure and enhance UX in mobile application design. It focuses on obtaining user behavior data from application logs and using deep neural networks to simulate user experience during app interactions. Kuang et al. \cite{Kuang} investigated the potential of proactive conversational assistants to aid UX professionals through automatic suggestions at three distinct times: before, in sync with, and after potential usability problems.

\subsection{Explainable AI (XAI)}

Explainable AI, or XAI refers to a set of techniques and methodologies to make the outputs of AI systems more understandable and interpretable by humans \cite{XuXai}. XAI is also a critical component in the development of AI-enabled UX tools. As outlined by Doshi-Velez and Kim \cite{Doshi-Velez}, the need for transparency and interpretability in AI systems is paramount, especially in user research contexts. 

\subsection{Platforms}

Several platforms have focused on developing user research by integrating AI-driven software for customer feedback analysis. Dovetail \cite{Dovetail} is a platform designed to assist teams in the analysis, organization, collaboration, and storage of user research data and insights. It allows researchers to tag and annotate qualitative data, integrate quantitative insights, and generate detailed reports to aid decision-making processes. Another platform is Maze \cite{Maze}, which focuses on product discovery by transforming prototypes into actionable insights. Maze facilitates the validation of ideas, testing of user flows, and collection of early feedback, enabling teams to optimize designs based on real user interactions. Similarly, LookBack \cite{Lookback} offers capabilities for a quick and thorough review of conversations, thereby reducing the workload associated with sorting and summarizing interview data. It supports live interviewing, remote usability testing, and self-guided testing that features automatic transcription and session recording to streamline user research methodologies.

However, the majority of the existing works have focused on developing software solutions that automate various aspects of the interview process, with an emphasis on post-interview analysis. These tools often aim to minimize or even replace the role of the human interviewer. In contrast, InsightPulse takes a different approach by integrating both hardware and software to \textbf{enhance}, rather than \textbf{replace}, the human interviewer. Unlike other systems, the InsightPulse hardware device provides real-time assistance during the interview, offering features such as automatic summaries and proactive follow-up question suggestions. This real-time support assists the interviewer to maintain focus and adapt to the conversation on the spot. In addition, the backend software system simplifies the post-interview analysis and research processes.

\section{Methodology}

\subsection{Overall System Architecture}

The architecture of InsightPulse integrates both hardware and software components to streamline the interview process through real-time support and automated data analysis. Fig. \ref{architecture} illustrates the overall system architecture of InsightPulse. The system comprises a microphone and a microprocessor for capturing and converting speech to text, the OpenAI Application Programming Interface (API) for natural language processing and analysis, and a web application interface managed by Streamlit.

The data flow begins with the microphone capturing the interview speech, which is then processed by the microprocessor to convert the audio into text. This text data is subsequently sent to the OpenAI API for analysis, which generates two outputs: (i) Output 1: Real-time Summary, and (ii) Output 2: Follow-up Question Suggestions. The two output modes can be accessed by the interviewer with a single tap for the summary or a double tap for follow-up questions on the device, which is displayed on the device screen.

The raw text data is also stored in a database, which can be accessed via the web application for post-interview analysis. This analysis includes identifying key discussion points, analyzing keyword frequency, conducting sentiment analysis, and generating thematic summaries.

\subsection{Hardware Design}

\begin{figure}[htbp]
\centerline{\includegraphics[width=8cm]{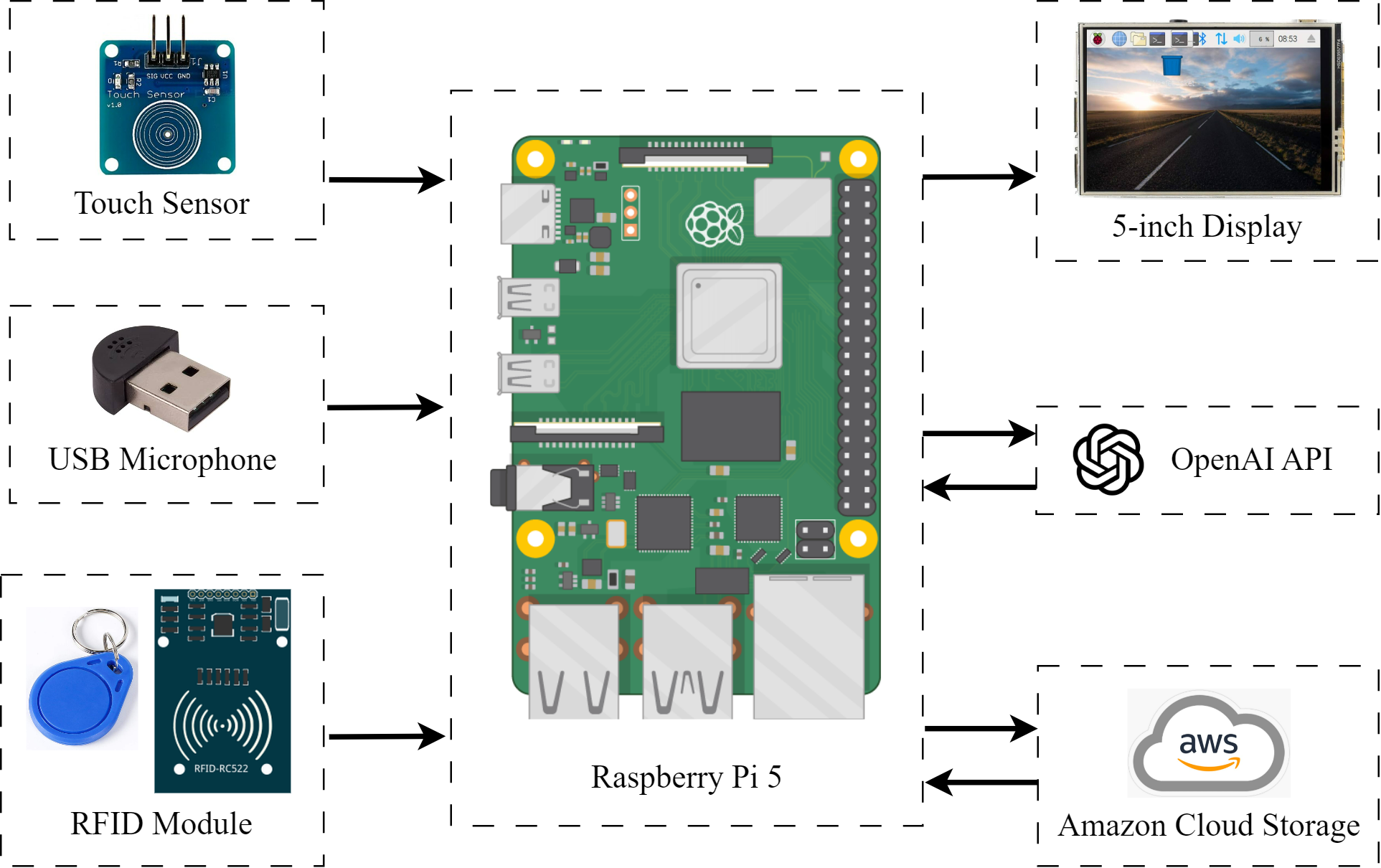}}
\caption{InsightPulse hardware setup.} \label{hardware}
\end{figure}

To ensure the protection of user privacy during the interview process, the interviewer needs to use a dedicated RC522 RFID-based control card to turn the device on. Once powered on, the interview content is captured by the microphone, and the Raspberry Pi 5 microprocessor records and converts the audio data to text. The text data is sent to the OpenAI API for detailed analysis, where it undergoes some preset questions to generate a response. The interview text data and the OpenAI-generated response are stored in Amazon Web Services (AWS) cloud storage. A 5-inch display shows key information in real-time during the interview. For example, when the interviewer taps the touch sensor once, the screen will display a key summary of the interview content so far, helping the interviewer understand the key insights up to that point. When the interviewer taps the touch sensor twice, the screen will display additional interview questions that can be further explored.

\begin{figure}[htbp]
\centerline{\includegraphics[width=8cm]{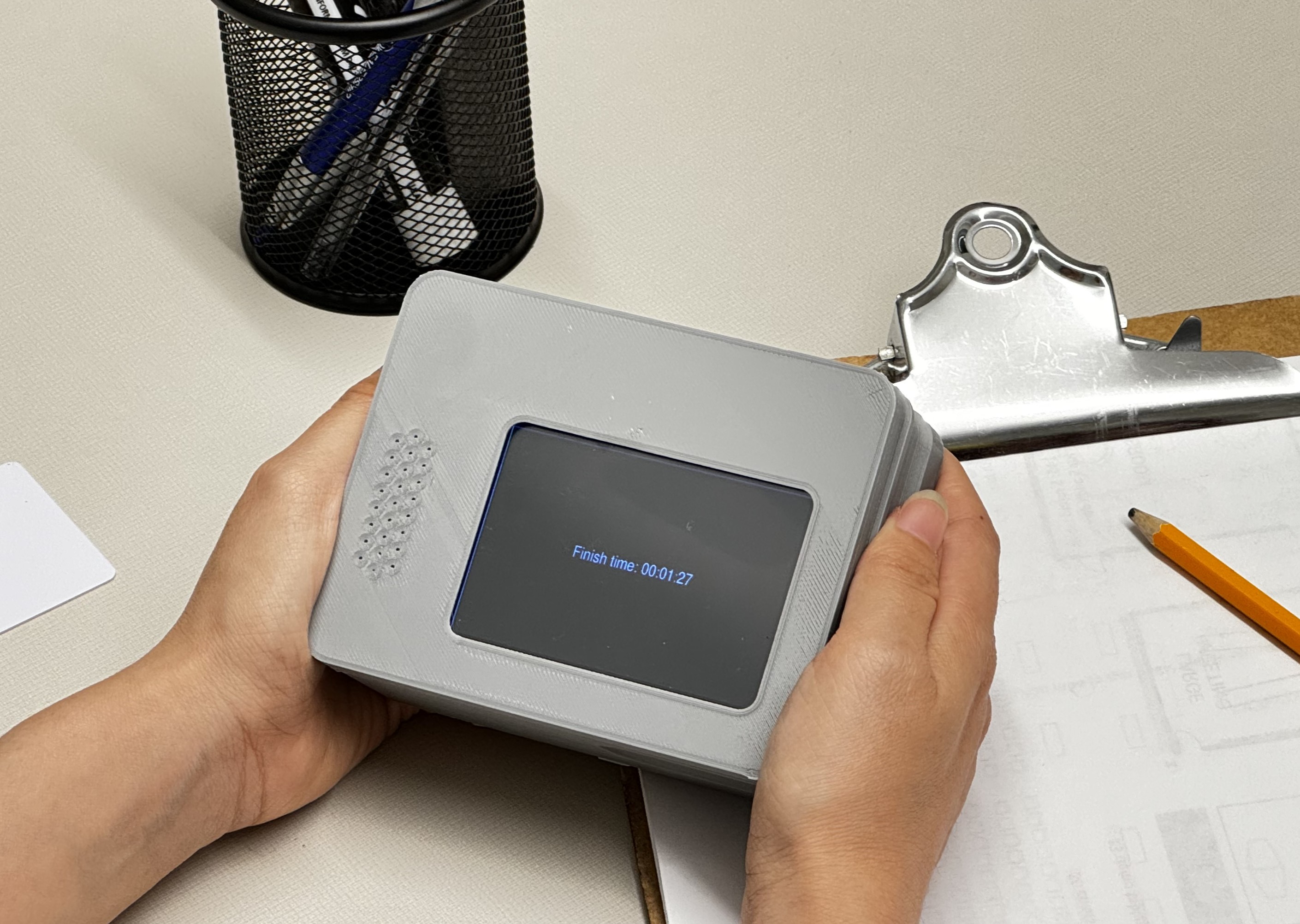}}
\caption{Assembled InsightPulse hardware device. Image captured during an ongoing UX interview test.} \label{assembled}
\end{figure}

To make the device portable, we are currently using a rechargeable 5000 Milliampere-hour (mAh) lithium-ion battery to power the device. Fig. \ref{hardware} illustrates the hardware setup of the InsightPulse device, and Fig. \ref{assembled} shows an assembled device during a real-world UX interview assistance test.

\subsection{Software Design}

The present software framework of InsightPulse is structured around (i) a Backend Analysis, and (ii) a Layered Software Architecture.

\subsubsection{Backend Analysis}

One of the most significant tasks of UX researchers is the organization and analysis of completed interview content. Summarizing key information and extracting insights from these interviews can be time-consuming and labor-intensive. This challenge is particularly pronounced in commercial organizations where multiple interviewers conduct numerous interviews. In such environments, user researchers often struggle to quickly comprehend the interviews conducted by their colleagues, leading to an overwhelming amount of data and content that needs to be processed \cite{Rowley}.

To address these challenges, we have developed a post-interview AI-assisted backend system with a user interface web application to categorize interview sessions and streamline the analysis process. The system extracts stored interview texts from the database, allowing researchers to view the original list of interviews and their associated details. Each interview detail page provides a comprehensive summary of the interview, enabling quick review and understanding.

For a more structured analysis, we have established six key tags to categorize the interview content at the sentence level. These are: (i) Needs and Expectations, (ii) Pain Points, (iii) Functionality and Features, (iv) Scenarios (When/ How/ Who/ Where/ Frequency), (v) Attitude (Positive/ Negative), and (vi) No Label. 
These tags allow user researchers to see the content aggregated according to these dimensions. These tags are customizable, allowing researchers to adapt the system to their specific research needs. For instance, researchers can identify which pain points were mentioned by users and what positive attitudes were expressed during the interviews. The tagging system enables researchers to understand the key takeaways and insights of an entire interview session without the need to analyze each interview individually. 


\subsubsection{Software Architecture}

\begin{figure}[htbp]
\centerline{\includegraphics[width=7 cm]{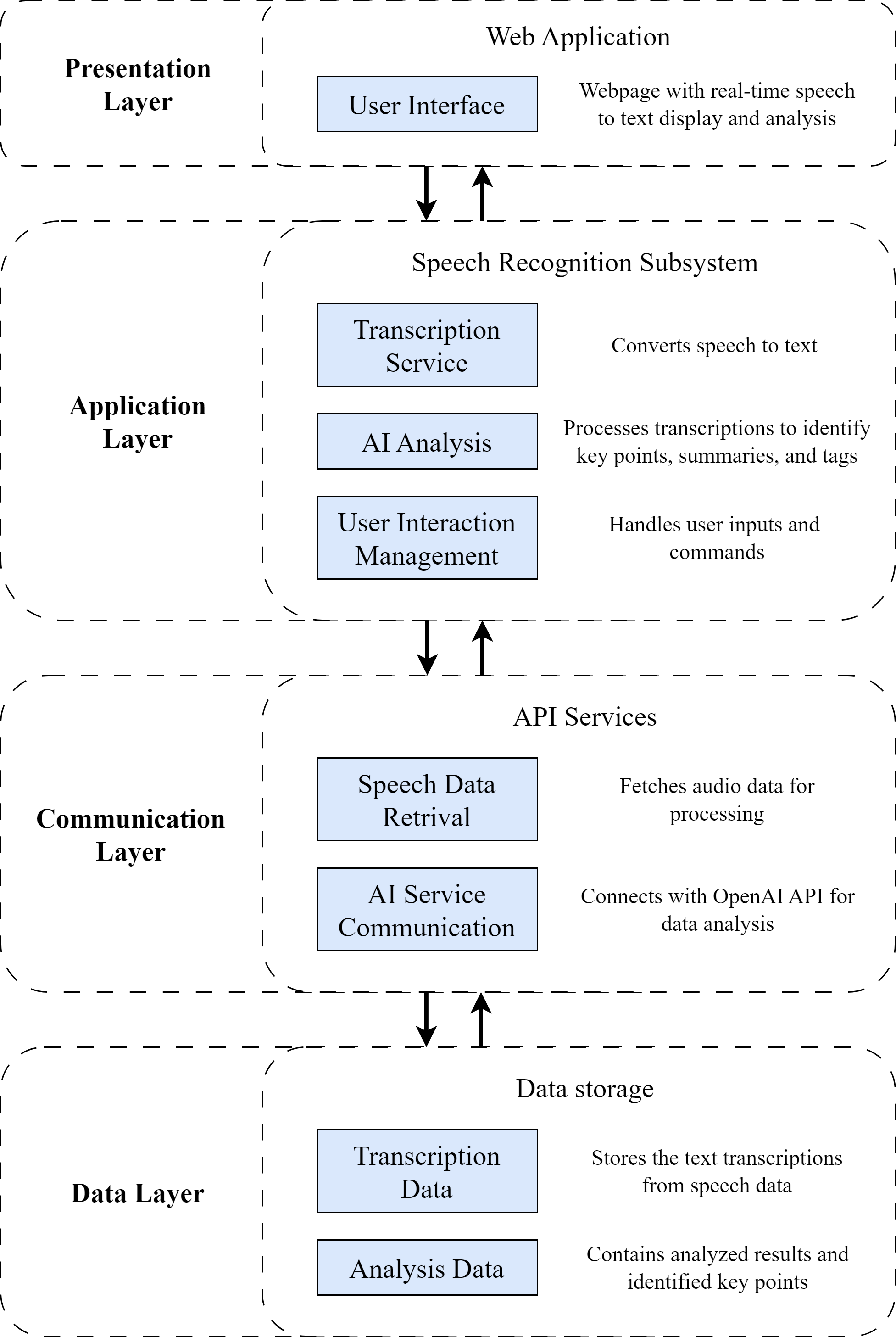}}
\caption{InsightPulse software architecture.}
\label{software}
\end{figure}

InsightPulse software system is designed using a multi-layered software architecture that ensures efficient processing, analysis, and interaction with speech data. The architecture consists of four primary layers: (i) Presentation Layer, (ii) Application Layer, (iii) Communication Layer, and (iv) Data Layer. Each layer is responsible for specific functions and interacts seamlessly with the other layers to provide a robust and user-friendly system. Fig. \ref{software} illustrates the layered software architecture of InsightPulse. 

\begin{figure*}[htbp]
\centerline{\includegraphics[width= 15cm]{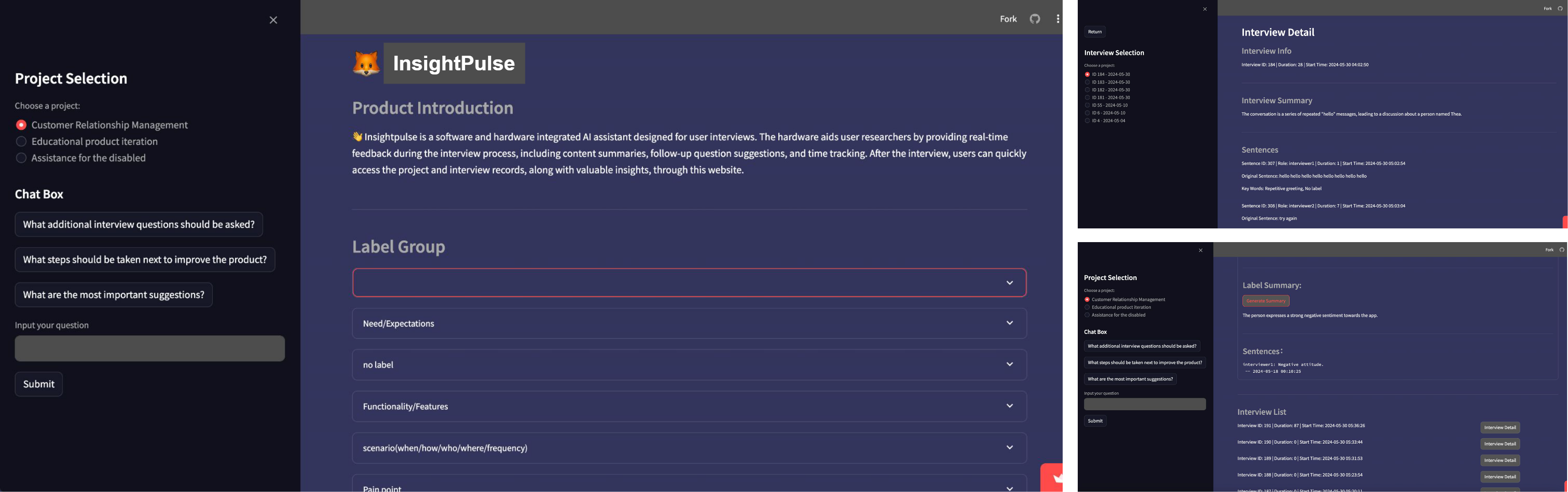}}
\caption{InsightPulse web application.}
\label{webapp}
\end{figure*}

The Presentation Layer is the user-facing part of the system comprising the web application. This layer provides the User Interface (UI) with a webpage that displays the real-time speech-to-text conversion and analysis results. The interface allows users to interact with the system and view transcriptions, summaries, and key points extracted from the speech data. Fig. \ref{webapp} shows screenshots of the web application that we are currently developing.

The Application Layer contains the core functionalities of InsightPulse speech recognition system. It is responsible for converting speech to text, analyzing the transcriptions, and managing user interactions. The Transcription Service within the Application Layer converts the incoming speech data into text format using speech recognition algorithms. The AI Analysis component processes the transcriptions to identify key points, generate summaries, and tag relevant sections. It leverages AI and natural language processing techniques to enhance the quality and relevance of the analysis. User Interaction Management handles user inputs and commands, ensuring that the system responds appropriately to user requests and interactions.

The Communication Layer manages the interaction between the application and external services. The Speech Data Retrieval component fetches the audio data that needs to be processed by the transcription service. It ensures that the audio data is correctly captured and transferred to the appropriate components for processing. The AI Service Communication component connects with external AI services, in our case the OpenAI API, to enhance data analysis and provide advanced features like natural language understanding and machine learning-based insights.

The Data Layer is responsible for storing and managing all data generated and used by the system. The Transcription Data component stores the text transcriptions generated from the speech data. It serves as a repository for all transcribed text, which can be retrieved for further analysis or review. The Analysis Data component stores the results of the AI analysis, including identified key points, summaries, and tags. It ensures that all analyzed data is systematically organized and available for user access through the presentation layer.

\subsection{Design Consideration}

The design of InsightPulse is driven by a comprehensive approach that prioritizes privacy, scalability, user-centricity, integration capability, and cost-effectiveness. The system accommodates the varying needs of different organizational sizes and project scopes, which allows for adaptation from small teams to large enterprises without disrupting existing workflows. The user interface of InsightPulse is designed to be customizable and adaptable to users with diverse technological proficiencies to reduce the learning curve for new users. 

\section{Cost and Testing}

\subsection{Cost}

The approximate cost of developing a single unit of the InsightPulse hardware and software system is detailed in Table \ref{cost}. The components are selected for their balance between performance and cost-effectiveness, thus ensuring the affordability and functionality of the system. The total hardware cost is approximately \$90.45. The API costs are calculated by estimating approximately 100 sets of interviews in one device, corresponding to the typical number of interviews conducted in a single UX research project.  This brings the total software cost to \$8.02, resulting in an estimated overall cost of \$98.47 for a single unit of InsightPulse system. These costs can be significantly reduced if the system is produced at scale.


\begin{table}[htbp]
    \centering
    \caption{Cost Estimation of Developing Single Unit of InsightPulse}
    \label{cost}
    \begin{tabular}{|c|c|c|c|c|}
    \hline
    \textbf{Component} & \textbf{Unit Price (USD)} & \textbf{Units} & \textbf{Total (USD)} \\
    \hline
    5-inch Screen & 17.00 & 1 & 17.00 \\
    \hline
    Raspberry Pi 5 & 70.00 & 1 & 70.00 \\
    \hline
    RFID Module & 0.95 & 1 & 0.95 \\
    \hline
    Touch Sensor & 0.50 & 1 & 0.50 \\
    \hline
    Enclosure & 2.00 & 1 & 2.00 \\
    \hline
    \multicolumn{3}{|r}{\textbf{Hardware Total =}} & \textbf{90.45} \\
    \hline
    HW API & 0.04 & $\approx{100}$ & 4.00 \\
    \hline
    SW API & 0.04 & $\approx{100}$ & 4.00 \\
    \hline
    Hosting (AWS) & 0.01 & 2 & 0.02 \\
    \hline
    \multicolumn{3}{|r}{\textbf{Software Total =}} & \textbf{8.02} \\
    \hline

    \multicolumn{3}{|r}{\textbf{Total Cost =}} & \textbf{98.47} \\
    
    \hline
    \end{tabular}
\end{table}

\subsection{Testing}

\begin{table*}[htbp]
\centering
\caption{Comparison of Traditional UX Interviews and InsightPulse-Assisted Interviews}
\label{comparison-table}
\begin{tabular}{|l|c|c|}
\hline
\textbf{Aspect} & \textbf{Traditional Method} & \textbf{InsightPulse-Assisted} \\ \hline
\textbf{Interviewer Interaction} & High engagement required; risk of bias & Automated consistent responses; reduced bias \\ \hline
\textbf{Respondent Comfort} & Dependent on interviewer skill; variable & Consistent neutral prompts \\ \hline
\textbf{Data Accuracy} & Dependent on transcription quality, not easily quantifiable & Quantifiable accuracy with automated transcription \\ \hline
\textbf{Preparation Time} & 6 hours & 2 minutes \\ \hline
\textbf{Interview Duration} & 1 hour per interview & 30-40 minutes per interview \\ \hline
\textbf{Post-Interview Processing} & 8 hours & 5 minutes \\ \hline
\textbf{Scalability} & Limited by interviewer availability & Easily scalable \\ \hline
\textbf{Flexibility in Adjustments} & Time-consuming mid-study adjustments & Easy dynamic adjustments \\ \hline
\textbf{Tool Requirements} & Multiple tools (Figma, Zoom, Google Docs, etc.) & Only InsightPulse hardware and web application \\ \hline
\textbf{Adaptability to Different Scenarios} & Significant adaptation needed & Adaptable with custom prompts \\ \hline
\textbf{Initial Setup Cost} & Moderate & Moderate \\ \hline
\textbf{Long-Term Maintenance Cost} & High ongoing costs & Low (mainly API usage costs) \\ \hline
\textbf{Overall Satisfaction} & 4.5/5 & 4.5/5 \\ \hline
\textbf{Insight Depth} & High but variable & High with consistent depth \\ \hline
\textbf{Categorization Quality} & 6 groups (manual categorization) & 8 groups (automated categorization) \\ \hline
\end{tabular}
\end{table*}

At present, the InsightPulse device is undergoing planned tests to ensure its reliability in real-world UX interviews. The functionalities and use cases have been detailed in the final project demonstration of the ``TECHIN 515: Hardware Software Laboratory 2" course offered by the Global Innovation Exchange (GIX) at the University of Washington, Seattle, WA, USA. 
As part of our primary test, a comparative study was performed where the same UX researcher conducted two rounds of interviews and subsequent post-interview analyses on the topic of Alzheimer's patient care--- (i) one using traditional methods and multiple interview tools, and (ii) the other using InsightPulse with real-time assistance. The test involved three UX researchers/ interviewers, each conducting five sets of interviews. The interviewers asked a combination of open-ended and structured questions to understand the key challenges and experiences associated with Alzheimer's patient care. The questions covered topics such as caregiver difficulties, communication with patients, and common healthcare practices. Table \ref{comparison-table} presents the insights derived from this comparative study based on the responses from the interviewers. The study reveals that InsightPulse significantly enhances the experience of the interviewer, particularly by reducing the \textbf{Preparation Time} (from 6 hours to 2 minutes) and \textbf{Post-interview Processing time} (from 8 hours to 5 minutes). In addition to the primary test, our ongoing testing process and plans involve:

\begin{itemize}
    \item \textit{Real-World UX Interview Questions:} The device is being tested with actual UX interview questions to evaluate how effectively it captures and processes responses. This includes checking the accuracy of real-time summaries, follow-up question suggestions, and thematic summaries.

    \item \textit{Relevance Verification:} We are verifying the relevance of the insights and follow-up questions suggested by the device with UX professionals. This test ensures that the responses align closely with the context and objective of the interviews.

   \item \textit{Comparative Analysis:} We plan to conduct multiple comparative analyses between traditional real-world interviews and post-interview analysis performed with and without InsightPulse assistance. After each session, the interviewer will rate the experience with a score.
   

\end{itemize}

Through these testing phases, we aim to refine InsightPulse and ensure that it meets the high standards required for effective UX research. The insights gained from real-world testing will be instrumental in optimizing the device's performance and enhancing its utility in diverse interview settings. 



\section{Limitations and Future Work}

While the InsightPulse system offers a solution to make the traditional UX interview process accessible, certain limitations need to be addressed as we improve the system. The current limitations and our future work to address them are as follows: 

\begin{itemize}
    \item \textit{Dependency on Third-Party AI Services:} The system currently relies on OpenAI's API for natural language processing, which raises concerns about privacy, potential ethical issues, and scalability. To address this, we are developing a comprehensive database of real-world UX interviews and analyses conducted by UX researchers. With this database, our objective is to design proprietary AI algorithms that improve data processing efficiency and reduce the dependency on third-party services.

    \item \textit{Onboard Computation:} The current version of InsightPulse does not include onboard computation which is essential in environments with limited internet connectivity or stringent data privacy requirements. Future iterations will explore more powerful local processing options of large language models on edge devices like Raspberry Pis \cite{RahmanQ} to reduce reliance on cloud services, improve latency, and enhance data privacy. 
    
    \item \textit{Possibility of Hallucinations in AI-Generated Summaries:} As with many large language models, there is a risk of hallucinations, where the AI may generate inaccurate or irrelevant information and affect the quality of insights derived from the interview. Future iterations of the InsightPulse system will include fine-tuning the developed AI models with domain-specific data and incorporating human verification to ensure the accuracy of real-time summaries and suggestions.

    \item \textit{Limited Dataset for Testing:} The current testing phase is based on a relatively limited set of interviews due to a lack of related public datasets. A more diverse test dataset is needed to fully assess the system's performance across different UX scenarios. In future work, we plan to expand the test data by incorporating a broader range of UX interview scenarios, encompassing a wider variety of interview types and participants to enable a more comprehensive evaluation. 
    
\end{itemize}

Our future work will focus on enhancing both the hardware device and the AI capabilities of InsightPulse. Continuous feedback from users will be essential to refine the user interface and the web application to make the system more intuitive and user-friendly.

\section{Conclusion}

The paper introduced the InsightPulse hardware and software system for UX research assistance. By offering real-time support, such as automatic summaries, proactive follow-up questions, and thematic analysis, InsightPulse aims to address common challenges that UX researchers face, including maintaining focus, managing interview duration, and simplifying post-interview analyses. The system's development and initial testing have demonstrated its potential to improve the efficiency and effectiveness of UX interviews. Through ongoing development and refinement, InsightPulse aspires to become an indispensable tool for UX interviewers and significantly enhance the quality and efficiency of UX research.


\bibliography{references}

\vspace{12pt}

\end{document}